# Silicon Nitride External Cavity Laser with Alignment Tolerant Multi-Mode RSOA-to-PIC Interface

Ibrahim Ghannam, Bin Shen, Florian Merget, and Jeremy Witzens

*Abstract*—We demonstrate an external cavity laser formed by combining a silicon nitride photonic integrated circuit with a reflective semiconductor optical amplifier. The laser uses an alignment tolerant edge coupler formed by a multi-mode waveguide splitter right at the edge of the silicon nitride chip that relaxes the required alignment to the III-V gain chip and equally splits the power among its two output waveguides. Both the ground and first order mode are excited in the coupler and reach the quadrature condition at the waveguide junction, ensuring equal power to be coupled to both. Two high-quality-factor ring resonators arranged in Vernier configuration close a Sagnac loop between the two waveguides. In addition to wideband frequency tuning, they result in a longer effective cavity length. The alignment tolerant coupler increases the alignment tolerance in the two directions parallel to the chip surface by a factor 3 relative to conventional edge couplers, making it ideal for gain chip integration via pick-and-place technology. Lasing is maintained in a misalignment range of ±6 µm in the direction along the edge of the chip. A Lorentzian laser linewidth of 42 kHz is achieved.

## I. INTRODUCTION

EXTERNAL cavity lasers (ECL) have been incorporated in systems servicing diverse applications such as coherent communications [1], spectroscopy and sensing [2], [3], swept-source coherence tomography [4] and length metrology [5]. The versatile application fields of ECLs are owed to their low linewidth and wide tuning ranges, but also to the possibility of maintaining compact sizes with complete or partial integration. Low linewidths make them suitable for applications that require long coherence lengths or low phase noise. Besides coherent communications with complex higher-order constellation diagrams and increased data rates, these also include quantum optics applications and optical atomic clocks [6]. Large tuning ranges make them attractive for applications such as swept-source coherence tomography and optical spectroscopy.

A great deal of interest has been given to the implementation of ECLs with a combination of III-V gain materials with a silicon photonic integrated circuit (PIC), to supplement the silicon platform with light emission functionality. While heterogeneous III-V on silicon integration has been pursued as a means to achieve dense integration and tight, lithography driven alignment tolerances [7], [8], much interest has also been given to the integration of off-the-shelf semiconductor optical amplifiers (SOAs) with silicon photonics PICs. Hybrid integration schemes can rely on micro-lenses or other interposed micro-optics [9] or on direct grating assisted [10] or butt-coupling [11] between flip-chipped components. While this enables the use of mature high-power and efficient reflective (R)SOAs and reduces the requirements for front-end-of-line (FEOL) PIC fabrication, it also results in substantial challenges related to assembly tolerances and internal losses at the gain chip to PIC interface.

There have been a number of successful demonstrations of hybrid-integrated ECLs, with past work focusing on reduction of the linewidth, increase of the tuning range, or both [9]-[11]. Compact laser sizes of a few millimeters or below allow for a broad range of applications and robust assemblies for use in demanding environments.

Here, in order to make such assemblies better manufacturable and thus better suited for mass-production, we focus on the problem of efficiently coupling light between the two elements, while relaxing alignment tolerances and showing that the other essential laser qualities, such as tuning range and linewidth, can be maintained. To address this, we make use of alignment tolerant couplers that we first demonstrated in silicon and optimized for edge [12] or surface [13] coupling. These relax the alignment tolerances by about a factor 3 in one direction compared to standard single mode coupling devices, at the price of requiring two on-chip output waveguides over which the power is equally split. This makes them ideal for applications such as parallel single mode transmitters [12] or passively biased Mach-Zehnder modulators [14], [15], but also tunable back-reflectors as required for ECLs and implemented by forming a Sagnac loop. First results for an ECL using silicon alignment tolerant grating couplers were reported in [16], but this device still required a pair of ball lenses assembled together with the RSOA over the chip in a cumbersome assembly. Moreover, it did not implement wideband tunability nor featured the low linewidths shown here and did thus only provide a first proof-of-principle of the utilized PIC topology.

This work was supported in part by the European Commission in the H2020 Framework Programme under Grant 688519.

I. Ghannam, B. Shen, F. Merget and J. Witzens are with the Institute of Integrated Photonics, RWTH Aachen University, 52074 Aachen, Germany (e-mail: ighannam@iph.rwth-aachen.de, bshen@iph.rwth-aachen.de, fmerget@iph.rwth-aachen.de, jwitzens@iph.rwth-aachen.de).

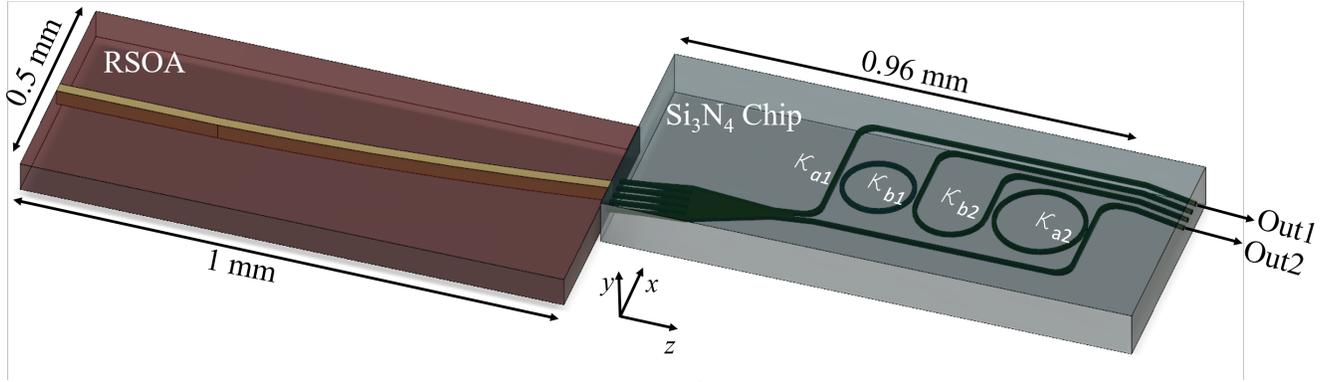

Fig. 1. Schematic diagram of the PIC and RSOA.

Here, we use an edge coupler configuration compatible with flip-chip integration technology and a pair of ring resonators in Vernier configuration for providing a wide, single mode tuning range. The PIC is implemented in a high-confinement silicon nitride (SiN) integrated waveguide platform [17] that provides both tight bends as well as reduced losses of ~0.1 dB/cm in the C-band, enabling large on-chip delays (implemented here with high quality (Q-)factor ring resonators) and thus substantial linewidth reduction. The SiN platform has a number of advantages when it comes to the implementation of ECLs, in particular a reduced thermo-optic coefficient increasing wavelength stability relative to silicon PICs [18], as well as low waveguide losses enabling long delays and narrow linewidths [19]. In low confinement platforms, it has yielded some of the best linewidths demonstrated to date for integrated solutions [20], [21]. Besides its desirable properties for ECL implementation, the SiN platform has also served for direct integration of rare earth based gain materials [22], [23].

Here, we focus on relaxing the required placement accuracy of the RSOA in a SiN based ECL. Compared to a conventional edge coupler, the required accuracy with which the III-V gain chip has to be placed in the two directions parallel to the surface of the PIC is relaxed by a factor 3, uniquely enabling assembly with pick-and-place flip-chip technology.

The rest of this paper is organized into three sections. In section two, we describe the silicon nitride PIC. In section three, we report the ECL characteristics, and finally, in section four, we conclude the paper and give an outlook on future work.

## II. PHOTONIC INTEGRATED CIRCUIT

Figure 1 shows a schematic of the PIC along with the RSOA and Fig. 2 a micrograph of the PIC. Light coupled from the RSOA is equally split between two waveguides at the output of the alignment tolerant edge coupler (ATEC), which are further coupled to each other through the two ring resonators. When the ring resonators' resonances are tuned to overlap with each other at a single wavelength, the two rings selectively close a Sagnac loop for that resonant wavelength. The rings are designed so that most of the light remains within the laser cavity. The coupling coefficients $\kappa_{a1}$ and $\kappa_{a2}$ describing the coupling strength between the main bus waveguides (connected to the ATEC) and the rings are chosen such that between 10% and 30% (depending on the design types described below) of the optical power is routed to the two chip/laser outputs labeled as Out 1 and Out 2 at the resonance wavelengths. The other coupling sections, with coefficients $\kappa_{b1}$ and $\kappa_{b2}$, connecting the two rings with each other via a third bus, are optimized so that all the light passing through this intermediate waveguide remains inside the laser cavity (i.e., the two unlabeled output ports between Out 1 and Out 2 were included as low reflection terminations and for monitoring purposes, but nominally carry zero output power). To reduce unwanted back-reflection into the PIC, the output edge couplers are slanted with a 15° angle relative to the normal to the interface, as parasitic back-reflections can lead to unstable operation or an increased linewidth. Compared to straight edge couplers with back-reflections extracted to be -14.5 dB from ripples in Fabry-Perot test structures implemented for that purpose, the waveguide-to-waveguide back-reflections of the slanted edge couplers were measured to be reduced to -19.5 dB. Slanting the edge couplers does not reduce the coupling efficiency to a lensed fiber to which light is coupled at the output of the ECL, so long as the latter is also aligned with a corresponding angle of 22° resulting from refraction.

The PIC was fabricated by LIGENTEC in their standard multi-project wafer (MPW) process with photonic structures fully etched in an 800 nm SiN film fabricated by high-quality low pressure chemical vapor deposition (LPCVD) and fully clad by SiO₂. In the following, the ATEC and the Vernier structure are described in more details.

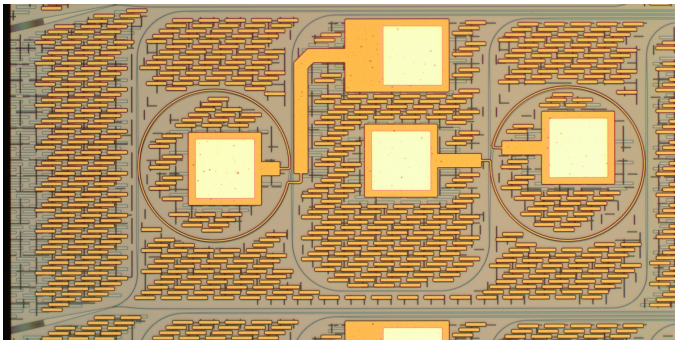

Fig. 2. Microscope image of the PIC. The output edge couplers are truncated on the right side.

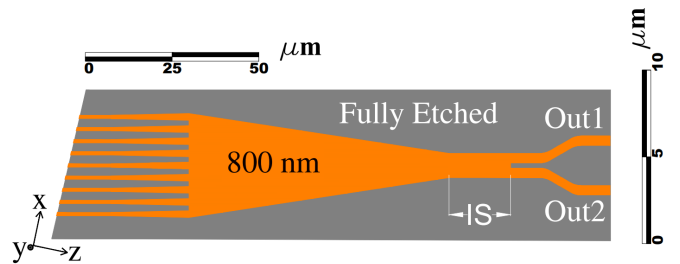

Fig. 3. Schematic of the ATEC. The scales in the horizontal and vertical directions are chosen to be different in order to facilitate visualization of the device (given its narrow aspect ratio), but it is otherwise drawn to scale. IS stands for interference section.

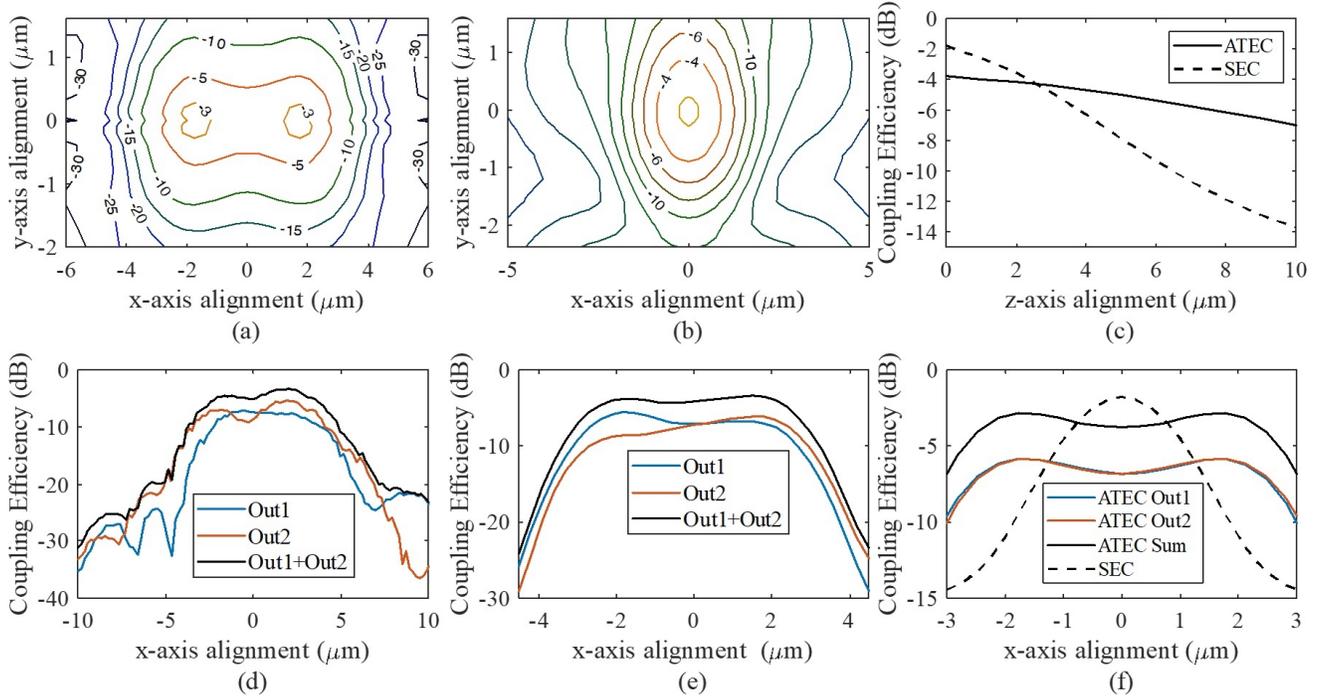

Fig. 4. Alignment dependent ILs for the ATEC and the SEC. Alignment dependent losses for (a) the ATEC (sum over both waveguides) and (b) the SEC for displacements in the xy plane parallel to the facet of the chip. (c) ILs for axial displacements along the z-direction (for centered xy alignment). (d) Measurement of ILs of an ATEC for displacements in the x-direction. Corresponding simulation results assuming (e) a 5° angular offset and (f) nominal angular alignment. All simulations were run using a Gaussian beam with a MFD of 2.5 µm matched to the MFD of the lensed fiber with which the measurements were done. The PIC-to-fiber output edge coupler losses were normalized out, as they are not part of the device characteristics.

### A. Alignment Tolerant Edge Couplers

Figure 3 shows the schematic of the ATEC. Light is first coupled into an array of 9 thin waveguides. These are 200 nm wide and spaced, center-to-center, by 700 nm. Since light exits the RSOA at an angle, also to reduce back-reflections, the waveguide array is also slanted by 13.1° to match the direction of propagation of the light after refraction. Together, the waveguides form, edge-to-edge, a 6 µm wide interface along the facet of the PIC and form the equivalent of a slab waveguide with weaker mode confinement in the vertical (y-)direction. The mode field diameter (MFD) of its supermodes, defined as the distance between the points at which the field intensity reaches $1/e^2$ of its maximum, is 2.5 µm in the y-direction and is closely matched to the vertical MFD of the RSOA, estimated from far field diffraction angles extracted from beam profile measurements, and to the MFD of the lensed fiber used for PIC characterization measurements in the following. The array of tips supports 4 transverse electric (TE) polarized supermodes with different numbers of in-plane lobes (along the x-direction), however, only the ground and first-order modes are being used, as the other modes are being filtered out by the downstream taper and interference section. The ground mode has an in-plane MFD of 4.5 µm, along the x-direction, also close to that of the RSOA but significantly larger than that of the lensed fiber used for PIC characterization. In either case, as the RSOA or lensed fiber are being displaced in the x-direction, the ground and first order supermodes are being excited with different amplitudes, but with the same phase (assuming the waist of the incoming beam is right at the interface).

The tips are then progressively tapered up from 200 nm to 400 nm, over a length of 30 µm, pulling the field into the SiN core, before being merged into a single slab. A second taper adiabatically reduces the slab width back to 1.4 µm, over a conservatively chosen length of 100 µm. The following interference section supports two TE-polarized modes, the ground and first order mode. The length of the multimode interface section is chosen such that the two lowest supermodes at the input interface are mapped to its two modes with a 90° relative phase shift at the output waveguide junction. Finally, the light is split and routed to two 550 nm wide single-mode output waveguides that are further tapered up to 600 nm.

This leads to the two modes always reaching the waveguide junction in quadrature, irrespective of the lateral (x-direction) displacement of the RSOA or lensed fiber at the input interface, suppressing interference and ensuring that equal power is coupled into both waveguides. This does not violate the reciprocity principle, as light is coupled to the two output waveguides with a RSOA-position-dependent relative phase. However, since the light is routed to a Sagnac loop thereafter, whose functionality does not suffer from this phase offset, it is irrelevant to the overall functionality of the chip. If the amplitude of the fields coupled into the two output waveguides of the ATEC are $a$ and $b$, the power coupled back by the Sagnac loop to the lensed fiber or RSOA is $4|ab|^2$ irrespectively of the relative phase of these two coefficients and is maximized for balanced output powers. This assumes that the loop itself is lossless. In practice, the insertion losses (ILs) of the Vernier structure have to be added to the overall loss budget.

Prior to conduction experiments with the RSOA, as reported in Section III, the PIC is characterized using a lensed fiber also at its input interface.

Corresponding simulation and measurement results are shown in Fig. 4. Simulated ILs are shown for different lensed fiber positions, with displacements in the xy-plane parallel to the facet of the chip assuming the input coupler to be the ATEC, Fig. 4(a), or a standard edge coupler (SEC), Fig. 4(b), assumed to be tapered down to a 200 nm tip width, as this was simulated to give the best insertion efficiency with the utilized lensed fiber. The -1 dB and -3 dB alignment tolerances of the ATEC are respectively 4 and 3 times better than for the SEC along the x-direction, as also seen in Fig. 4(f). These are defined as the range of allowable displacements to maintain the ILs within 1 dB or 3 dB of their optimum (i.e., for the ATEC the reference is conservatively taken at x=±1.7 μm). They are ±0.6 μm (SEC, -1 dB), ±2.4 μm (ATEC, -1 dB), ±1 μm (SEC, -3 dB) and ±2.9 μm (ATEC, -3 dB). However, there is also a 1 dB penalty in the peak coupling efficiency of the ATEC (-2.8 dB vs. -1.8 dB) resulting from the multiple transitions in the structure. It should be noted though that in a conventional ECL PIC consisting in a SEC followed by a separate 1-by-2 splitter, the latter would also result in some amount of additional ILs.

In the vertical y-direction, the ATEC does not improve the alignment tolerance, since in both cases it results from overlap integrals between field profiles with closed to matched MFD. The -3 dB alignment tolerance is actually slightly worse for the ATEC (±0.7 μm vs. ±1 μm for the SEC), as a consequence of it being referenced to the IL optimum at x=±1.7 μm but being taken at x=0 μm. This is however acceptable for the application pursued here, as the ATEC is meant to facilitate flip-chip integration for which the accuracy in x and z is determined by the placement accuracy, but the vertical alignment is defined by mechanical contacts, for example between the III-V chip and pedestals formed in a common substrate [11] or in the PIC [24].

In the z-direction, the other direction in the plane of the chips, there is also a significant improvement in alignment tolerance, as seen in Fig 4(c). In order to maintain the ILs within 3 dB of their optimum, the beam waist of the lensed fiber emission can be in a range between 0 μm to 3 μm from the edge of the PIC for the SEC. For the ATEC, this range is increased from 0 μm to 7.1 μm. This is due to the supermodes of the coupled waveguide tips having a much wider width than the SEC mode, resulting in reduced diffraction (considering the reciprocal coupling problem). Thus, the ATEC provides relaxed and acceptable alignment tolerances in both in-plane directions compared to the capabilities of off-the-shelf pick-and-place processes, as the overall required alignment accuracy of ±3 μm is well in range of post-cure placement accuracies. These simulation results are summarized in Table 1.

Measurements of the x-alignment dependent ATEC ILs are shown in Fig. 4(d) for both output waveguides. Losses occurring at the output of the PIC, at the interface to the lensed fibers picking up the transmitted signal, are normalized out as they are extrinsic to the device. The best coupling efficiency that was measured is -3.3 dB, 0.5 dB below the simulated number. There is a slight imbalance between the two outputs, which is attributed to a small deviation of the interference section length from the quadrature condition. We also observe a slight asymmetry in the alignment dependent data, i.e., moving from –x to +x does not result in exactly permutating the waveguide dependent ILs. This could be attributed to a small deviation in the incidence angle of the input light beam (with respect to the z-axis) from the designed for angle, as a consequence of experimental conditions in the test setup holding the input lensed fiber. This is illustrated by a simulation with an input angle offset by 5° from nominal (Fig. 4(e)). For comparison, the nominal case is also shown in Fig. 4(f). In this case, the two outputs are perfectly symmetric and equal for all x-displacement values.

### B. Vernier Structure

The use of a single or of multiple ring resonators is a common practice in integrated ECL design for providing wavelength selectivity. Ring resonators can be interposed in the optical path between the chip interface and a reflective element [11], used directly as a wavelength selective reflector [19], or as part of a Sagnac loop [1], [7]-[10]. The use of multiple rings in Vernier configuration for wideband laser tuning [1], [7], [11] was also demonstrated for other wavelength ranges, e.g., in the O-band [8].

For the ECL at hand, a Vernier structure consisting of two rings is implemented, that are each individually tunable with thermal tuners over an entire free spectral range (FSR) with an 80 mA current range and below 200 mW of dissipated power (per ring). Due to the low loss of the SiN waveguide platform, it is possible to design ring resonators with very high quality (Q-)factors. On the other hand, the bending radii (and accordingly, the radius of the ring resonators) must remain larger, in comparison to Si, due to the reduced index contrast between SiN and SiO₂. This, in turn, leads to a small FSR, which can impair the functionality of the ECL by causing mode hopping since a large number of resonances then fall within the gain bandwidth of the RSOA if a single ring is used.

TABLE I
SIMULATED ILS AND ALIGNMENT TOLERANCES

|  | Insertion Losses | 1-dB Tol. in x | 3-dB Tol. in x | 3-dB Tol. in y | 3-dB Tol. in z |
|---|---|---|---|---|---|
| SEC | 1.8 dB | ±0.6 μm | ±1 μm | ±1 μm | 0-3 μm |
| ATEC | 2.8 dB | ±2.4 μm | ±2.9 μm | ±0.7 μm | 0-7.1 μm |

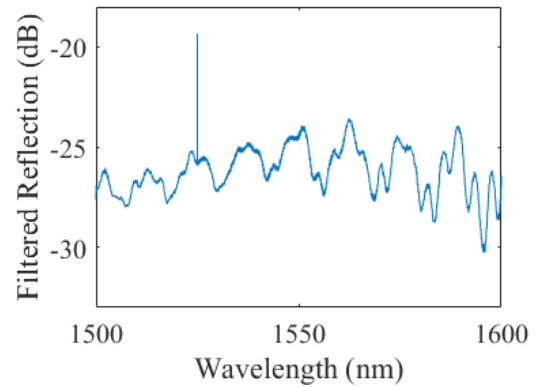

Fig. 5. Wavelength selective back reflection from the PIC (Design 2) measured with a circulator. A lensed fiber with a 2.5 μm MFD was used. High losses are due to a large separation between the lensed fiber and the chip in this experiment, that served to characterize the extinction ratio.

Two ECL designs were implemented on-chip with different ring Q-factors, as determined by the coupling strengths to the bus waveguides. For each ECL design, two rings with the same loaded Q-factor and slightly different radii of 115 μm and 117 μm are used.

The high Q-factor effectively creates a long cavity length and therefore removes the need to use long delay lines to reduce the phase noise and the linewidth. The effective length of a ring at resonance, $L_{eff}$, can be estimated as

$$L_{eff} = \frac{2Q_L c_0}{\omega n_g} \tag{1}$$

where $Q_L$ is the loaded quality factor, $c_0$ the speed of light in vacuum, $\omega$ the angular resonance frequency, and $n_g$ the group index. However, as the targeted Q-factor is increased and the coupling strengths consequently reduced, the ILs resulting from waveguide losses and excess junction losses inside the rings also go up. The two designs are summarized in Table 2.

TABLE II
ECL DESIGNS

| | Loaded Q-factor | Eff. length (per ring) | Insertion loss (per ring) | Through power[1] | Coupling coeffs.[2] ($\kappa_a, \kappa_b$) |
|---|---|---|---|---|---|
| Design 1 | 215 K | 5.1 cm | 1 dB | 10% | 0.11, 0.13 |
| Design 2 | 600 K | 14.2 cm | 2 dB | 30% | 0.06, 0.07 |

[1]Power remaining in the main bus waveguides connected to the ATEC, on resonance.

[2]Amplitude coupling coefficients.

The effective FSR of the Vernier structure can be roughly estimated as [25]

$$FSR_V = \frac{FSR_1 \cdot FSR_2}{|FSR_1 - FSR_2|} \tag{2}$$

with $FSR_1$ and $FSR_2$ the FSRs of the two rings equal to 1.6 nm and 1.62 nm, resulting in an effective FSR of 130 nm larger than the gain bandwidth of the utilized RSOAs (see Section III). To test the Vernier structure of Design 2, the heaters of the two rings are set to obtain overlapping resonances at 1525 nm. Light is coupled from the lensed fiber to the ATEC and the back-reflection recorded by means of a circulator. The resulting reflection spectrum is shown in Fig. 5. Since the position of the lensed fiber was not carefully optimized in this experiment and kept at a bigger distance from the PIC, absolute values are not representative. However, this data serves to verify the selectivity of the back-reflector. A single resonance remains in the range of 1500 nm to 1600 nm accessible by our test equipment, with all other resonances staying suppressed. Moreover, an off-resonance extinction ratio of 6 dB is obtained around the selected wavelength, that remains better that 4 dB over the entire recorded spectrum. Based on the independent characterization of the Vernier structure and of the ATEC, an on-resonance back-reflection coefficient of ~8 dB is expected fiber-to-fiber.

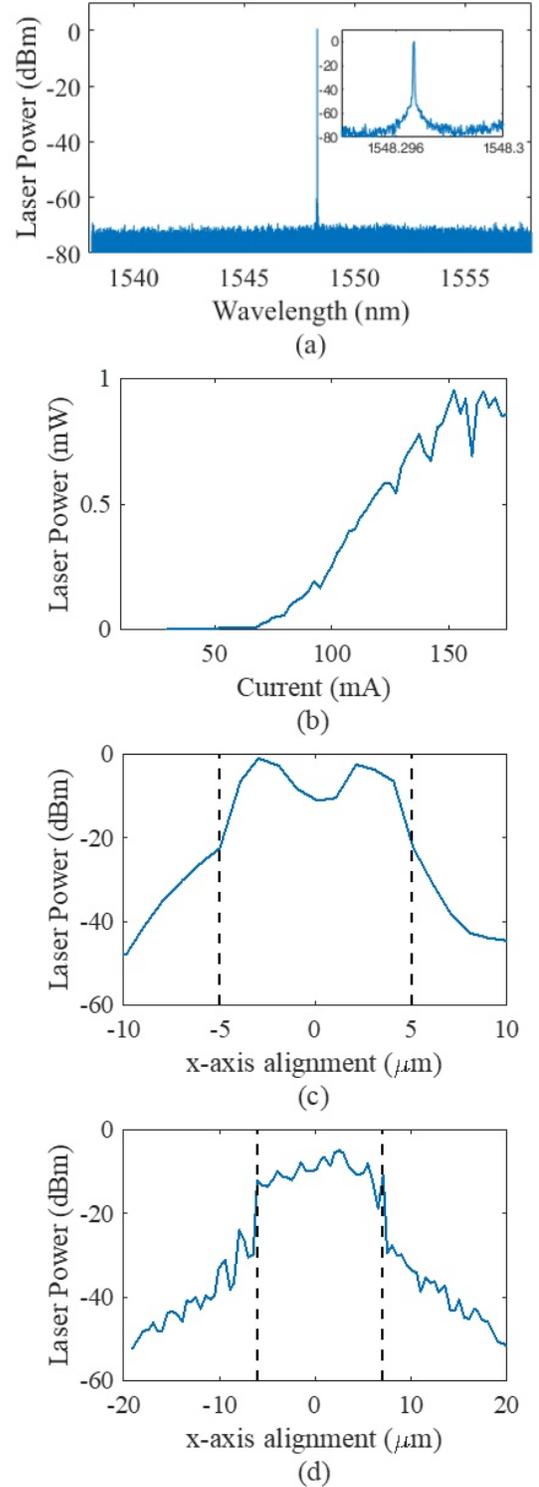

Fig. 6. (a) Laser spectrum for Design 2 operated with the C-band RSOA and an injection current of 200 mA, recorded with a resolution bandwidth (RBW) of 5 MHz. Power is indicated in dBm, since the entire laser line falls within the RBW and recorded power levels are directly plotted instead of being divided by it. (b) Laser output as a function of current with the Vernier structure tuned to 1530 nm. The threshold current is 70 mA. (c) Output power as a function of x-axis alignment. The Vernier structure was tuned to 1530 nm and the injection current set to 275 mA. (d) Output power as a function of x-axis alignment for Design 1 with the L-band RSOA. The Vernier structure was tuned to 1605 nm and the injection current set to 200 mA. In both cases, dashed lines indicate the range in which lasing occurs, marked by a sharp increase in the output power.

## III. EXTERNAL CAVITY LASER

After characterizing the PIC, the lensed fiber was replaced by an RSOA and a series of experiments were performed to characterize the ECL. We use commercial RSOAs operating in the C- and L-bands. The main specifications of the RSOAs are summarized in Table 3. The light beam exits the RSOA with a nominal angle of 19.5° relative to the surface normal of the output facet, in order to reduce internal reflections. Lasing spectra were recorded for both Vernier structure designs.

TABLE III
RSOA CHARACTERISTICS

|  | C-Band RSOA | L-Band RSOA |
|---|---|---|
| Center Wavelength | 1518 nm | 1572 nm |
| Vertical FWHM | 28.1° | 28.5° |
| Lateral FWHM | 15.7° | 14.0° |
| Current to reach transparency | 60 mA | 60 mA |
| Bandwidth (3dB) at 300 mA | 97.5 nm | 96.5 nm |
| Back Facet reflectivity | 90% | 90% |
| Front Facet reflectivity | < 0.01% | < 0.01% |
| Saturation power at 300 mA | ~60 mW | ~60 mW |

An example of the recorded optical spectrum, for a Design 2 device operated with the C-band SOA and an injection current of 200 mA, is shown in Fig. 6(a). The side-mode suppression ratio is above 73 dB at 1548 nm with the measurement limited by the noise floor of the spectrum analyzer. To confirm single mode operation, the spectrum was recorded over the entire range of the high-resolution optical spectrum analyzer (1520 nm to 1629 nm) and this was the only laser mode that was found. Figure 6(b) shows the optical power versus applied RSOA current (LI-curve) with the Vernier structure tuned to

1530 nm, close to the gain maximum of the RSOA. The onset of lasing is at an RSOA injection current of 70 mA, just slightly above the 60 mA required to obtain positive gain from the RSOAs (Table 3). Power levels reported in this section (Figs. 6(b)-6(d)) correspond to the power coupled to a lensed fiber at one of the output ports of the PIC (Out 1). Recorded power levels of ~1 mW are in the expected range, given the ~60 mW saturation power of the RSOA, the 3 dB losses at the RSOA to PIC interface, the 5-10 dB on-resonance extinction of the rings (the rest is being coupled back), the 3 dB PIC-to-fiber outcoupling efficiency of the SECs (that is experimentally slightly worse than simulated values), and the power being split over the two output waveguides.

The alignment tolerance between RSOA and PIC was tested by moving the RSOA along the x-direction parallel to the chip facet, with the Vernier structure on the PIC tuned to a given wavelength. The system's output power for Design 2 operated with the C-band SOA, as above, a 275 mA injection current and the Vernier structure tuned to 1530 nm is shown in Fig. 6(c). Lasing was obtained in a range of lateral displacements of ±5 μm, with boundaries shown by dashed lines in the figure. At the edges of this misalignment range, the output power sharply drops as the losses become larger than the small signal gain of the RSOA, lasing action ceases, and only amplified spontaneous emission (ASE) remains. Within the ±5 μm misalignment range, the characteristic shape of the ATEC can be recognized. However, the dependency of the laser output power on misalignment is much more pronounced than the misalignment dependent ILs recorded during the passive characterization measurements (Fig. 4).

This is a consequence of the laser output power depending nonlinearly on internal cavity losses close to threshold. This can be exemplified by describing the RSOA gain saturation with a simple model as $G = G_0/\left(1 + P_{in}/P_{sat,in}\right)$, with $G_0$ the small signal (round-trip) gain, $P_{in}$ the power at the input of the RSOA and $P_{sat,in}$ its 3-dB gain compression input saturation power. This results in $P_{in} = (G_0R - 1)P_{sat,in}$ with $R$ (the power) reflection coefficient of the PIC and $P_{out} = (G_0R - 1) P_{sat,in}/R$, with $P_{out}$ the power at the output of the RSOA. The power coupled to the PIC thus scales as $T(G_0R - 1) P_{sat,in}/R$, with $T$ the power coupling coefficient between the RSOA and the PIC. In the limit where $G_0R >> 1$, the laser output power is simply $TG_0P_{sat,in}$ and scales with the ILs of the ATEC. However, as $G_0R$ approaches 1, $T(G_0R - 1) P_{sat,in}/R$ depends strongly on $R$, that in turn scales as $T^2$. Below threshold, coherent emission collapses and the RSOA only emits ASE. The power generated by the device then also scales simply as $T$, albeit with a much lower proportionality factor.

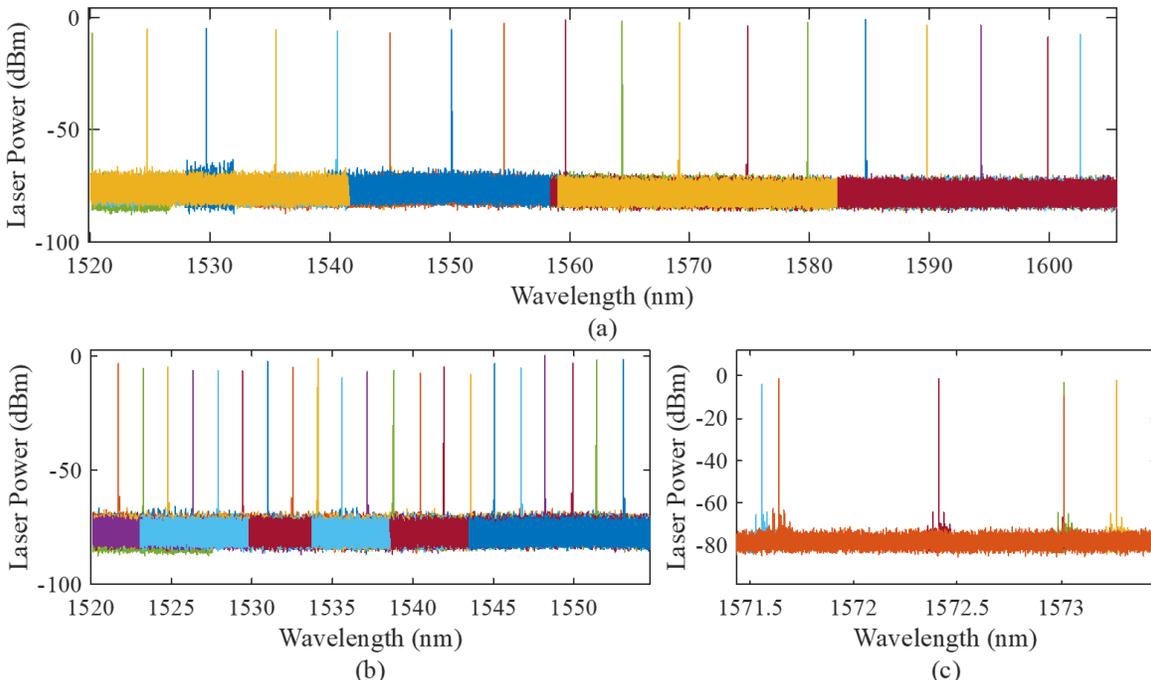

Fig. 7. (a) Complete tuning range of Design 1 operated with the C-band RSOA. The dataset is limited by the spectrometer range below 1520 nm. The injection current was set to 150 mA for wavelengths below 1570 nm and to 200 mA for longer wavelengths. (b) By changing the tuning current of one of the ring resonators' thermal tuners, it is possible to change the lasing wavelength by a whole FSR and (c) by tuning the thermal tuners of both rings the laser wavelength can be swept within a single FSR.

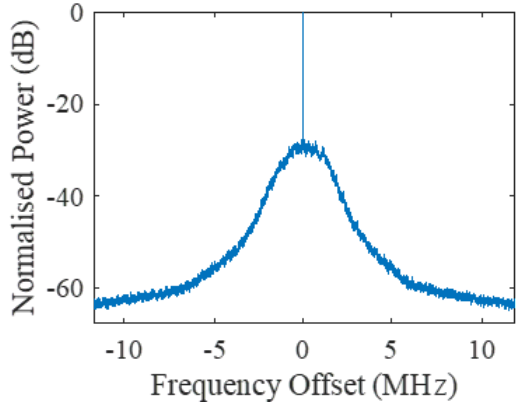

Fig. 8. Spectrum recorded from the delayed self-heterodyne linewidth measurement with Device 2 and the L-band RSOA. The injection current was set to 200 mA and the Vernier structure tuned to 1584 nm. The resolution bandwidth of the electrical spectrum analyzer was set to 10 kHz.

Fig. 6(d) shows a second measurement done for a Design 1 device operated with the L-band RSOA, an injection current of 200 mA, and the Vernier structure tuned to 1605 nm. The ECL was found to lase in a slightly wider range of lateral displacements covering ±6 μm. Also, the laser emission as a function of lateral displacements can be seen to be much more flattop in this dataset. This is somewhat surprising, as the injection current here was lower and the RSOAs, other than their shifted emission spectra, have very similar properties. Since the previously discussed dataset was obtained at a larger injection current and presumably a larger small signal gain, one would have expected a smaller rather than a larger sensitivity on coupling losses for the latter. It should be noted, however, that a number of experimental factors play a role on the laser characteristics recorded in individual datasets. In particular, for the dataset shown in Fig. 6(c), the RSOA remained at a significant distance, larger than 10 μm, from the PIC, as opposed to Fig. 6(d), for which the laser to PIC distance was more aggressively optimized (accepting the risk of potentially crashing the RSOA in the utilized setup). Further aspects such as small errors in angular alignment or a slight error in the length of the interference section in the different devices leading to a small offset from the quadrature condition may have further played a role here. These should however be straightforwardly addressed in a production environment in which placement accuracies much better than the 5º angular offset assumed in Fig. 4(e) as well as reproducible fabrication of optimized devices would be achievable (see below for an analysis of required fabrication tolerances).

Moreover, the ECL described here does not have a separate phase shifter to align the Fabry-Perot resonances resulting from the overall cavity formed by the RSOA and the PIC with the resonances of the rings. Given the 1 mm RSOA length and the equivalent delay lengths of the rings, we estimate the FSR of the Fabry-Perot resonances to be 1.4 GHz for Design 1 and 0.52 GHz for Design 2. The full width at half maximum (FWHM) of the rings, on the other hand, are 0.9 GHz and 0.32 GHz, and are below the corresponding Fabry-Perot FSR. Given that moving to the edge of the FWHM results in a 3-dB drop in input to drop port coupling efficiency per ring (i.e., 6 dB total), it can be seen that there can be a significant spread in internal laser cavity losses between experiments, as well as outcoupling coefficients, potentially leading to some spread in the recorded data. However, we did not experimentally observe any evidence of this being a significant problem, as stable lasing was obtained once the ring resonances were tuned to coincide, irrespectively of e.g. small variations in the RSOA to PIC distance influencing the spectral positions of Fabry-Perot resonances. Given the relatively high Q-factor of the utilized rings, frequency pulling might have played a role in facilitating alignment of the laser wavelength with the ring resonances [26], [27].

The next set of experiments aimed at characterizing the tunability of the ECL and were carried out with a Design 1 device and the C-band RSOA. By tuning the resonances of the two ring resonators on the PIC, a large tuning range measured to span over 83 nm was achieved. Measurements were limited by the range of our high-resolution spectrum analyzer at the lower wavelength boundary (1520 nm) and lasing was limited by the tapering off of the C-band RSOA gain in the higher wavelength range (lasing stopped a few nm above 1600 nm). Figure 7(a) shows the whole tuning range of the ECL (just selected lines are shown for clarity, so that the output power levels can be seen). Since the gain of the C-band RSOA is lower at the longer wavelengths, the injection current was increased for wavelengths above 1570 nm (150 mA below 1570 nm and 200 mA thereafter). By tuning one of the ring resonators while keeping the current applied to the thermal tuner of the second ring constant, it is possible to hop from one ring resonance to the next and thus change the lasing wavelength by one ring FSR (Fig. 7(b)). Furthermore, it is possible to tune the lasing wavelength within a single FSR of an individual ring resonator by changing the current applied to both rings (Fig. 7(c)).

The range of achievable lasing wavelengths is expected to be limited primarily by the gain spectrum of the RSOA, with the bandwidth of the ATEC playing a secondary role. Simulations show that an ATEC with an interference section length optimized to obtain the targeted quadrature condition and thus balanced waveguide output power levels at 1550 nm reaches an output imbalance of 3 dB at 1500 nm and 1600 nm. For this level of imbalancing, the reflection of the Sagnac loop into the RSOA mode, $4|ab|^2$ as discussed above, drops by only 0.5 dB. Moreover, the bandwidth of the ATEC can be improved by reducing its length [12], which ought to be possible here as the adiabatic transitions have been conservatively designed.

We also assessed the tolerance of the ATEC to manufacturing tolerances and found that increasing the width of all features by 20 nm over its entire length resulted in an imbalance of 4 dB, corresponding to the reflection dropping by 0.9 dB.

Finally, the linewidth of the Design 2 laser with L-band SOA, at an injection current of 200 mA and the Vernier structure tuned to 1584 nm, was estimated using the delayed self-heterodyne (DSH) measurement technique using a 6.5 km fiber delay. Figure 8 shows the result of the measurement. We used the model from [28] with a Voigt fit to extract a Lorentzian linewidth of 42 kHz with a 1.2 MHz Gaussian component.

## IV. CONCLUSION

In this work, we demonstrated an ECL with an alignment tolerant interface between the gain chip and the PIC. It allows for easier coupling and promises higher yield in high volume manufacturing. The alignment tolerance of the alignment tolerant edge coupler is three times better than that of a standard edge coupler in the x-direction, parallel to the edge of the chip, and improved by over a factor 2X in the axial z-direction, away from the edge of the chip.

High confinement silicon nitride waveguides, with low loss and medium mode confinement, allow for maintaining a compact ECL size. At the same time, it is possible to achieve very good results in terms of tunability, output power, and linewidth.

Lasing was maintained for x-axis displacements in a range of ±6 μm. Operated with a single C-band RSOA, the laser can be tuned in a range exceeding 1520 nm to 1603 nm, limited on the lower wavelength side by spectral range of our test equipment. A Lorentzian linewidth of 42 kHz was measured. Ongoing work aims at further reducing this linewidth by using even higher quality factors as well as antireflection coatings to reduce parasitic effects.

The main challenges seen while using the alignment tolerant couplers in a laboratory environment were proper angular alignment between the two chips and meeting the required quadrature condition inside the alignment tolerant edge coupler that ensures maximum alignment tolerance. These challenges should be overcome with controlled, automated assembly in a manufacturing environment.